\documentclass[prc,aps,twocolumn,showpacs,preprintnumbers,nofootinbib]{revtex4-1}

\usepackage{amsmath}
\usepackage{graphicx}
\usepackage{dcolumn}
\usepackage{bm}
\usepackage{slashed}
\usepackage{amssymb}
\usepackage{xcolor}
\usepackage{verbatim}
\usepackage{todonotes}
\usepackage{grffile}
\usepackage{soul}

\begin{document}

\author{Thorben Graf}
\affiliation{SUBATECH, UMR 6457, Universit́e de Nantes, Ecole des Mines de Nantes, IN2P3/CNRS. 4 rue Alfred Kastler, 44307 Nantes cedex 3, Franc}
\author{Jan Steinheimer}
\affiliation{Frankfurt Institute for Advanced Studies, Frankfurt am Main, Germany}
\author{Christoph Herold}
\affiliation{School of Physics, Suranaree University of Technology, 111 University Avenue, Nakhon Ratchasima 30000, Thailand}
\author{Marcus Bleicher}
\affiliation{Frankfurt Institute for Advanced Studies, Frankfurt am Main, Germany}

\title{Testing Charm Quark Equilibration in Ultra-High Energy Heavy Ion Collisions with Fluctuations}

  \begin{abstract}
Recent lattice QCD data on higher order susceptibilities of Charm quarks provide the opportunity to explore Charm quark equilibration in the early quark gluon plasma (QGP) phase. Here, we propose to use the lattice data on second and fourth order net Charm susceptibilities to infer the Charm quark equilibration temperature and the corresponding volume, in the early QGP stage, via a combined analysis of experimentally measured multiplicity fluctuations. Furthermore, the first perturbative results for the second and fourth order Charm quark susceptibilities and their ratio are presented.
\end{abstract}

\maketitle

\section{Motivation}
Heavy ion collisions serve as a tool to explore the properties of Quantum Chromo Dynamics (QCD) under extreme conditions. Experimental heavy ion programs are running at various collider facilities around the globe. The current high energy frontier is explored in Au+Au reactions at a center-of-mass energy of $\sqrt{s_{\mathrm{NN}}}=$200 GeV at the Relativistic Heavy Ion Collider (RHC) at Brookhaven, USA and at the Large Hadron Collider (LHC) at CERN with Pb+Pb reactions of up to $\sqrt{s_{\mathrm{NN}}}=$5.02 TeV. 

A description of the hadron multiplicities in terms of a statistical emission from a thermalized grand canonical hadron resonance gas at a temperature of 150-160 MeV has been rather successful for the majority of explored strange and non-strange hadron species.
Furthermore, fluctuations of conserved charges \cite{Ding:2015fca,Borsanyi:2014tda,Karsch:2015nqx} have been introduced as a novel tool to explore the chemical freeze-out conditions of the hadronic matter \cite{Ratti:2014jra,Bluhm:2014wha,Alba:2014eba}. The use of fluctuations (i.e. higher moments of the distribution related to the susceptibilities) are proposed as an alternative window to fix the temperature and chemical potential of the system at the chemical decoupling surface in contrast to the usual use of mean multiplicities or multiplicity ratios.
However, high statistics data samples from different experiments, e.g. recent ALICE data on (anti)-proton production, are currently challenging this straightforward interpretation (see discussion in \cite{Steinheimer:2016cir,Steinheimer:2012rd,Andronic:2012dm,Becattini:2012xb,Pan:2014caa}), highlighting the importance of the hadronic rescattering on the final observable hadron yields and multiplicity distributions.
  
In contrast to the light flavor hadrons, the Charm quark yield is treated as a conserved quantity throughout most of the systems evolution and is assumed not to change within the created quark gluon plasma or the following hadronic re-scattering. One usually assumes that Charm quarks are produced by initial high energy perturbative processes and are later distributed among the statistically available hadron species \cite{Andronic:2003zv}. However, if the initial temperature is sufficiently high, thermal Charm quark production in the very early phase of the collision may become feasible.

In this work we want to exploit the unique features of Charm production to gain insights into the early stage of the reaction, i.e. in the state of QGP matter before the hadronization. Comparing the calculated net-Charm susceptibilities \cite{Petreczky:2009cr} and experimental data, this may allow to gain insights into the initial temperatures reached in the reaction, complementary to direct photon measurements. We exemplify this idea on the Charm quark thermalization that might be explorable at current and future accelerators. Besides the established calculations of the Charm fluctuations on the lattice, we use a perturbative framework including bare quark mass dependence at next-to-leading order, presented in Refs.~\cite{Graf:2015tda,Graf:2016mzv}, to calculate the first results for second and fourth order Charm quark susceptibilities and their ratio with weak coupling techniques.

\section{Potential Charm quark equilibration}
Up to LHC energies, e.g. at RHIC, charm production generally proceeds in a perturbative process via the fusion of gluons from the initial nuclei. At RHIC, e.g. the initial temperatures are certainly too low to allow for thermal Charm production. This means that Charm production is not linked to the thermal properties of the medium and may only serve as kind of tomographic probe at these energies. At LHC energies and beyond the situation changes. Here the system created in central heavy ion collisions can reach initial temperatures of the order of 0.6-1 GeV. If the temperature during the system's evolution is comparable to the Charm quark mass, one would expect that these quarks reach their thermal equilibrium abundances sufficiently fast. It is not clear a priori, if this is the case in heavy ion collisions, which makes Charm quark thermalization an interesting topic for experimental investigations.

There are two central issues involved. The first is the problem of relaxation time scales. On the one hand Bodeker and Laine \cite{Bodeker:2012gs} estimated the Charm equilibration rate, using weak coupling, and found a very large relaxation time for Charm. On the other hand Zhou et al. \cite{Zhou:2016wbo}
found a relatively small relaxation time based on kinetic theory (which is also weak coupling inspired). In fact in Ref. \cite{Zhou:2016wbo} (see \cite{Zhang:2007dm} for the initial pioneering studies) it was suggested that thermal Charm quark production is the dominant process for Charm production in Pb+Pb collisions at $\sqrt s=39$ TeV, i.e. at the Future Circular Collider (FCC), while at current LHC energies it is already of substantial importance.

The second issue is that these weak coupling arguments, valid close to equilibrium,
may not be relevant at all at very high energy heavy ion collisions.
The dominant scale is the gluon saturation scale $Q_s$ which is
clearly above the Charm quark mass. One has a situation
where equilibrated matter that contains
light quarks and gluons as well as Charm quarks is created at a time
scale of the order of $1/Qs$. Then the initial temperature
will determine the Charm quark number. This is qualitative very different from the above weak coupling picture, where one first has thermalized gluons and light quarks and from them one creates the Charm quarks.

Therefore, at initial QGP temperatures, exceeding several hundred MeV, Charm quark equilibration is possible and should be addressed experimentally. This will allow direct access to the temperatures reached within the early QGP and may also contribute to the recent discussion about a flavour dependent decoupling temperatures. Such a scenario was put forward in ~\cite{Bellwied:2013cta}, where it was suggested that the hadron gas description of the lattice data on susceptibilities hints at two different (flavour dependent) freeze-out temperatures and it was speculated that a similar or even larger difference might be present in the charm sector. However, further studies suggest that in the charm sector the main difference between the hadron gas description and the lattice data is due to the lack of (up to now experimentally unobserved) charmed resonances in the hadron gas \cite{Bazavov:2014yba,Bazavov:2013dta}. It was suggested that the inclusion of charmed quark model states in the hadron gas also leads to a good description of the charm yields at higher temperatures without the assumption of a flavour dependent hadronization temperature. 
\begin{figure}[t]
\includegraphics[width=0.5\textwidth]{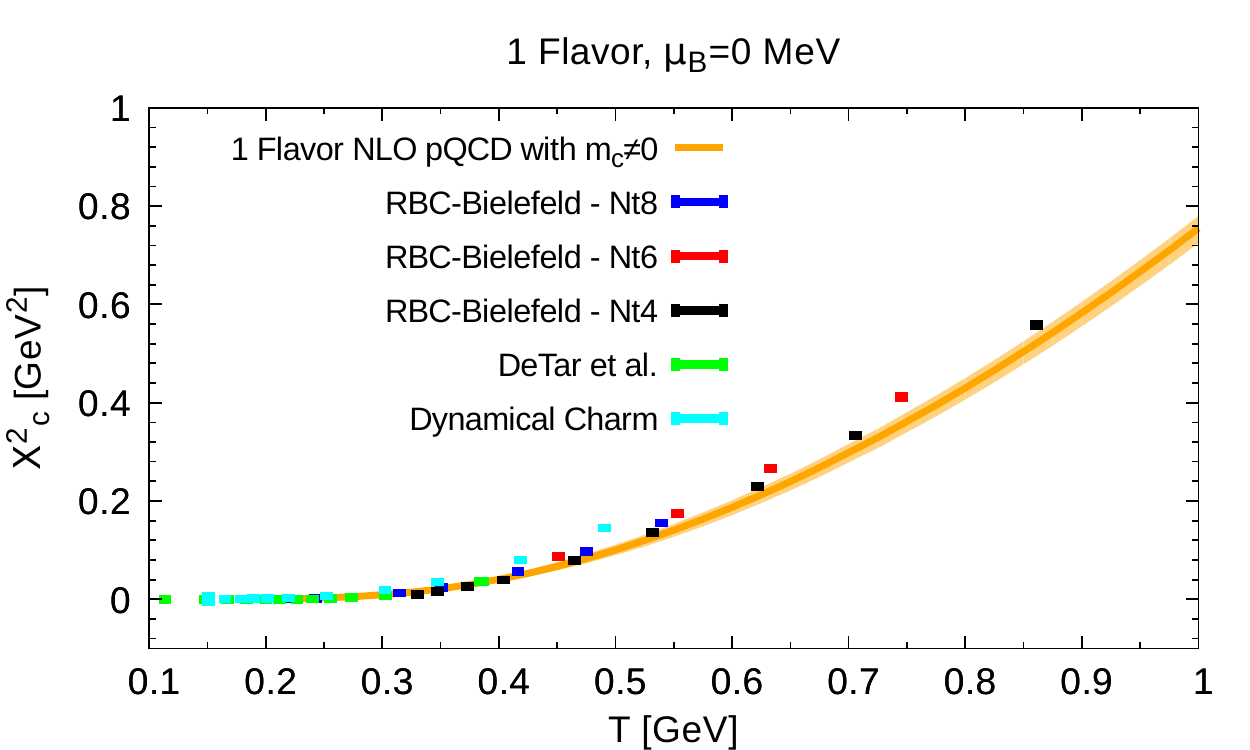}
\caption{\label{fig:susc2cc}(Color online) Second order Charm quark susceptibilities from our pQCD NLO framework (full line) in comparison to lattice QCD calculations (symbols) by the RBC-Bielefeld collaboration in quenched approximation for the charm \cite{Petreczky:2009cr}, from DeTar et al. in quenched approximation for the charm \cite{DeTar:2010xm} and from Borsanyi et al. for dynamical charm quarks \cite{Borsanyi:2012vn}.}
\end{figure} 
\begin{figure}[t]
\includegraphics[width=0.5\textwidth]{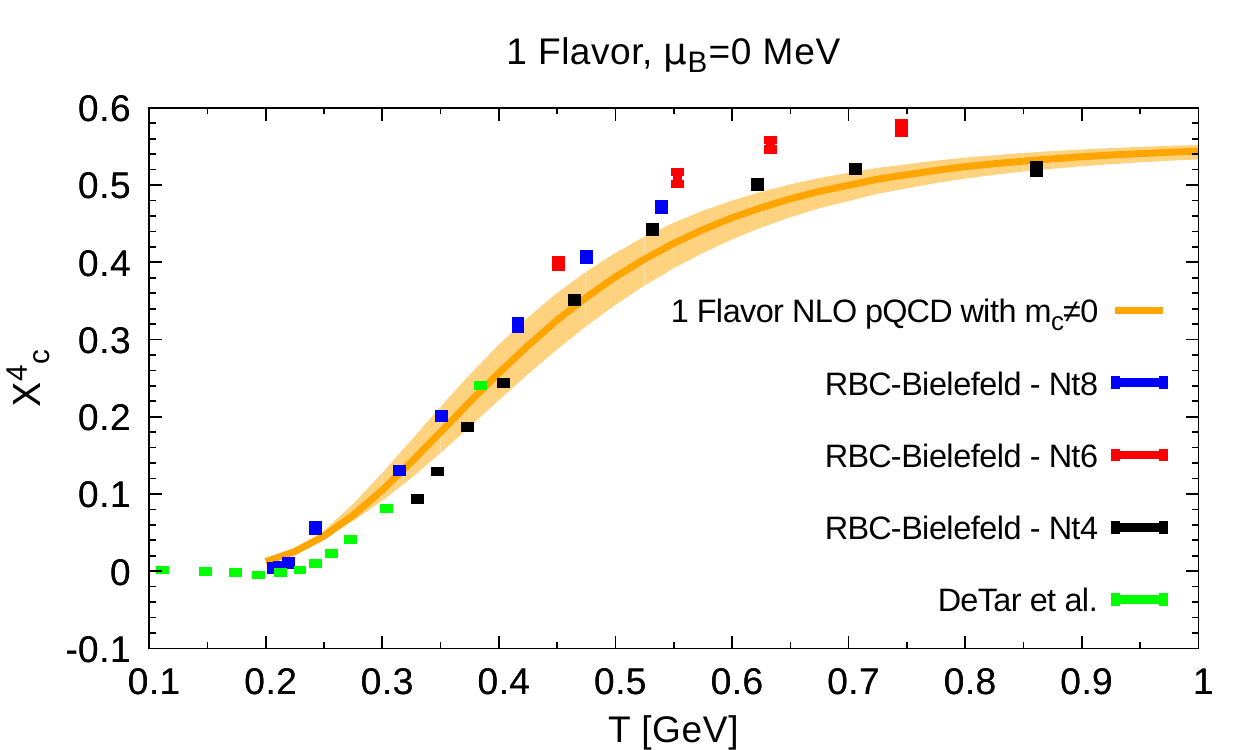}
\caption{\label{fig:susc4c} (Color online) Fourth order Charm quark susceptibilities from our pQCD NLO framework (full line) in comparison to lattice QCD calculations (symbols) by the RBC-Bielefeld collaboration in quenched approximation for the charm \cite{Petreczky:2009cr}, from DeTar et al. in quenched approximation for the charm \cite{DeTar:2010xm}.}
\end{figure} 

\section{Measuring the Charm quark equilibration temperature}
In this paper, we propose to perform a similar study as suggested in Ref. \cite{Alba:2014eba} for the strange and light quarks for Charm quarks.  The (quenched) lattice QCD data, in line with our pQCD approach allows to explore higher order fluctuations to serve as an experimental probe of Charm equilibration in the plasma stage of the reaction. Of course the extraction of the decoupling temperatures and volumes of Charm quarks relies on the  quality of the theoretical input and needs to be taken with a grain of salt as long as unquenched lattice QCD data at high temperatures are not available. Nevertheless, the general idea of the extraction of decoupling temperatures and volumes as presented below is independent on the details of the theoretical input.  
\begin{figure}[t]
\includegraphics[width=0.5\textwidth]{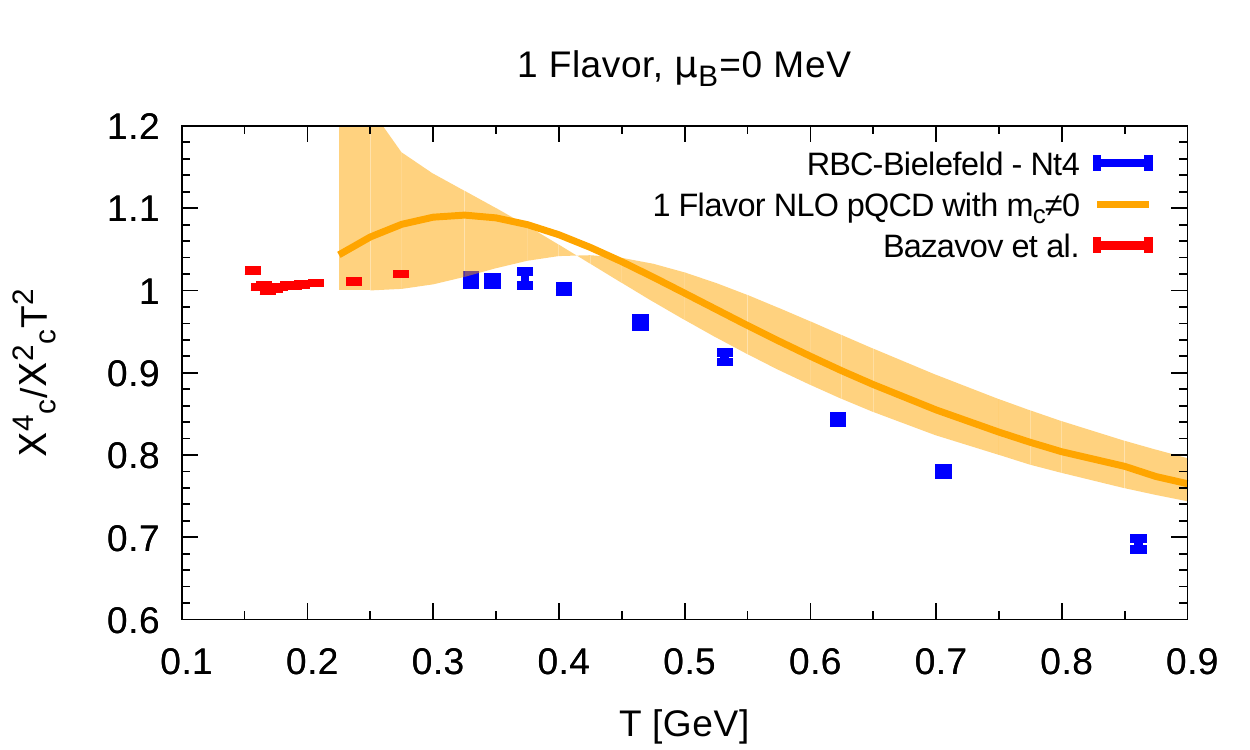}
\caption{\label{fig:suscratio42charm} (Color online) Charm quark susceptibilities ratios from our pQCD NLO framework and extracted from lattice QCD calculations by the RBC-Bielefeld collaboration in quenched approximation for the charm \cite{Petreczky:2009cr} and from Bazazov et al. \cite{Bazavov:2014yba}.}
\end{figure} 

The idea is that at high temperatures Charm quarks do reach chemical equilibrium during the system's evolution. If this is the case, then the fluctuations, e.g. the second ($\langle(\delta N_{c-\bar c})^2\rangle /\chi^2_c = T V$) and the fourth moment ($(\chi^4_c)/(\chi^2_c/T^2) = \kappa\sigma^2$) of the distribution (related to the susceptibilities taken from lattice QCD calculations) allow to extract the volume $V$ and the temperature $T$ of the system at the decoupling temperature of the Charm quarks, using the measured data on the kurtosis $\kappa$ and the variance $\sigma$ to extract $T$ and $V$. Generally the decoupling of a particle from a system is governed by two quantities, namely the expansion rate $\Gamma_{\rm exp}$ and the scattering rate $\Gamma_{\rm scat} (\sigma_i)$, where $\sigma_i$ is the inelastic cross section for the production of particle type $i$ . When the expansion rate exceeds the inelastic scattering rate the chemical reactions cease and the particle multiplicity distributions become independent of time (apart from decays, that are not relevant for Charm quarks in this scenario). In a simplified picture, one may assume that the condition $\Gamma_{\rm exp} \geq \Gamma_{\rm scat} (\sigma_i)$ coincides with a hyper-surface of constant chemical freeze temperature for the particle species $i$. In many studies this hyper-surface is for hadron states characterized by the so-called hadron-chemical freeze-out temperature $T_{\rm ch}\sim 150-160$ MeV. In a more realistic scenario, different inelastic processes have different cross sections (possibly dependent on quark and flavor content), which results in an extended emission process of light hadrons, suggesting a more differential point of view and a sequential freeze-out of hadrons \cite{Knoll:2008sc,Bellwied:2013cta,Noronha-Hostler:2014aia,Becattini:2014hla}. 

\begin{figure}[t]
\includegraphics[width=0.5\textwidth]{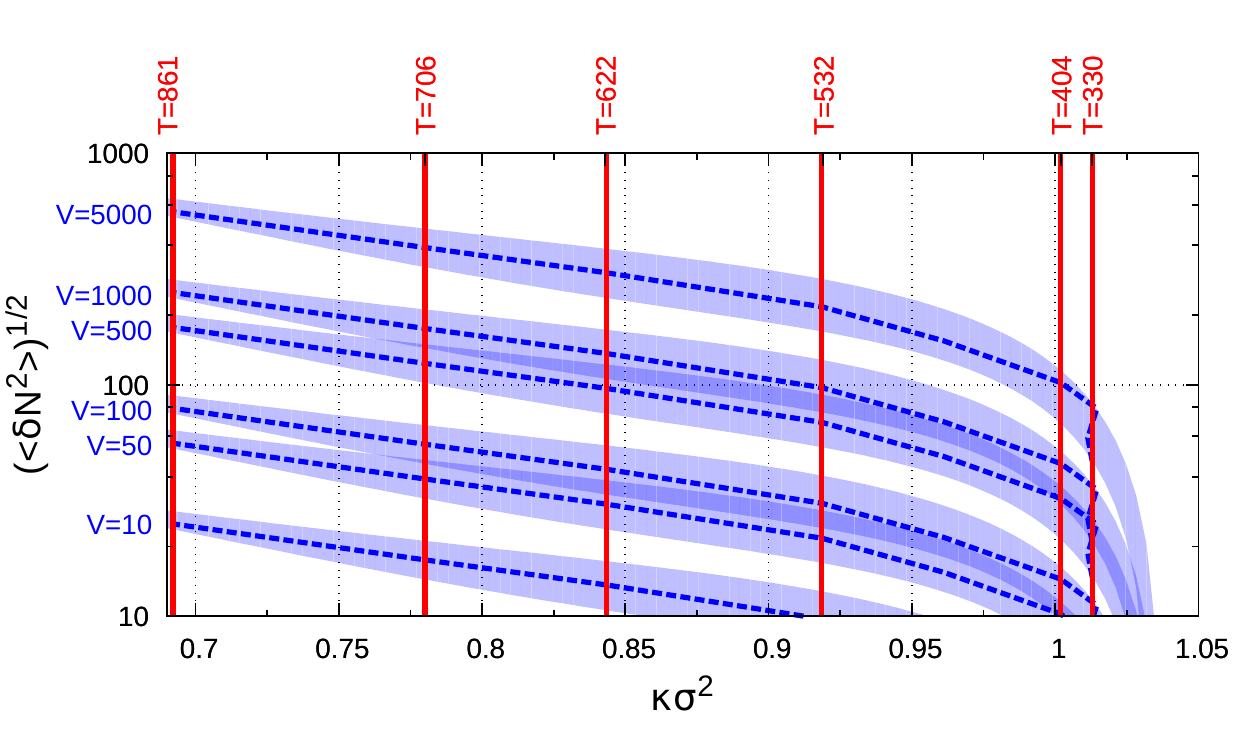}
\caption{\label{fig:ExPlot} Fluctuation of the square root of net Charm quark number $\sqrt{\langle(\delta N_{c-\bar c})^2\rangle}$ over higher moments $\kappa\sigma^2$, where lines of constant temperatures (solid) in $\text{MeV}$ and lines of constant volumes (dashed) in $\text{fm}^3$ are included, extracted from lattice QCD calculations by the RBC-Bielefeld collaboration \cite{Petreczky:2009cr}. To provide an estimate for the uncertainty of the results, we use the errors of the pQCD calculations and show them as bands.}
\end{figure} 

For Charm quarks, in contrast to light hadrons, the equilibration temperature is not related to hadronization but is expected to be on the order of a few hundred MeV. At a plasma temperature below 400-500 MeV, the dominant Charm production channel in the QGP, $gg\rightarrow c\bar{c}$, drops rapidly and Charm production (and annihilation) ceases \cite{Zhang:2007dm}. I.e., the Charm quark yield is fixed at the corresponding temperature and subsequent process will only alter the momentum distribution. If Charm quarks do indeed chemically equilibrate in the partonic phase, then the measurement of the multiplicity distribution, i.e. especially the second and higher moments of the net-Charm distribution of Charm quarks, allows for a comparison to the lattice data on Charm quark susceptibilities. The Charm susceptibility is defined as
\begin{equation}
  \left.\chi_c^{i}(T)\equiv\frac{\partial^{i}p(T,\vec{\mu})}{\partial\mu^i_c}\right|_{\vec{\mu}=0}.
\end{equation}

How can the Charm quarks fluctuation be extracted from experimental data? As discussed above the Charm quark distribution becomes frozen below the Charm decoupling temperature. The Charm quarks then proceed towards the hadronization hypersurface and form hadrons. Hadron formation proceeds either via fragmentation or via parton recombination with a light quark \cite{Prino:2016cni}. Both hadronization processes do not change the number Charm quarks, because (I)  the probability to create a new Charm quark pair in the fragmentation process is very small and (II) the annihilation probability of $c+\bar c \rightarrow xy$ ($x,y,$ not being charmed particles) is very small. Additionally one might have Charm quark diffusion towards unmeasured rapidity regions, e.g. due to interactions during the partonic stage or by thermal smearing. Both effects are however strongly suppressed for heavy particles. Thus, the finally observable D and $\bar{\rm D}$ yields (and other charmed hadrons, e.g. $\Lambda_c$) can be seen as a good proxy for the totally produced charm yield. Experimentally the measurement of D-mesons is straight forward, e.g. by employing a heavy flavor tracker.

In Fig.~\ref{fig:susc2cc} we compare our perturbative results for the second order Charm quark susceptibility with lattice QCD data in the quenched approximation for charm from ~\cite{Petreczky:2009cr,Bazavov:2014yba} and with dynamical charm quarks from Ref. \cite{Borsanyi:2012vn}. Unfortunately dynamical Charm quark data only exists up to temperature around 500 MeV. Up to this temperature region, the effect of dynamical charm is moderate and the quenched approximation provides a reasonable approximation as was shown in \cite{Borsanyi:2012vn}. Due to the lack of data it is a priori not clear how much the inclusion of dynamical Charm quarks will influence the charm quark susceptibilities at higher temperatures. Therefore, we rely for the quantitative estimates on the assumption that the quenched approximation is still reasonably good at higher energies. For the present setting we conclude that the quenched lattice QCD calculations and the perturbative QCD NLO result are apparently in good agreement, nevertheless it seems that the NLO pQCD calculation is slightly lower than the lattice data (especially when comparing to the last data point for dynamical quarks).  For the fourth order Charm quark susceptibility we can observe a similar behaviour, see Fig.~\ref{fig:susc4c}. Here we compare to the data from ~\cite{Petreczky:2009cr,DeTar:2010xm}, both in quenched approximation for charm. We can see that at mass scales of one GeV, which is the case for Charm at low temperatures, the impact of the quark mass dominates over the higher order corrections. Finally, we are interested in the ratio of these susceptibilities of different order, since this ratio can be related to the central moments kurtosis $\kappa$ and the variance $\sigma$ of the (measured) net Charm event-by-event multiplicity distribution as $(\chi^4_c)/(\chi^2_c/T^2) = \kappa\sigma^2$. The lattice QCD and perturbative QCD NLO results for the ratio are illustrated in Fig.~\ref{fig:suscratio42charm}. The perturbative band reproduces the trend of the lattice simulations correctly and only for high temperatures there is an increased deviation. 

A measurement of $\kappa\sigma^2$ (see e.g. Refs.~\cite{Adamczyk:2013dal,Adamczyk:2014fia} for measurements of higher moments of net-proton multiplicity distributions by STAR) of the net Charm number does therefore allow to extract the temperature of the Charm quark decoupling volume. With the known temperature, the second order susceptibility then allows to extract the volume in addition via $\langle(\delta N_{c-\bar c})^2\rangle /\chi^2_c = T V$. Thus, we have two equations
\begin{equation}
\langle(\delta N_{c-\bar c})^2\rangle /\chi^2_c = T V, \quad{\rm and}\quad  \chi^4_c/(\chi^2_c/T^2) = \kappa\sigma^2\quad,
\end{equation}
and two unknowns, $T,V$. This system of equations can then be solved to extract combinations of $T,V$ that are compatible with the experimentally measured values for $\kappa\sigma^2$ and $\langle(\delta N_{c-\bar c})^2\rangle$ of the charm quarks. The fluctuations of the net Charm quark numbers can be observed in the final Charm hadron distributions in a sufficiently sized rapidity window. Due to the large mass of the Charm quark, dilution effects, as maybe present for light quarks, can be expected to be negligible. As a result one obtains the volume $V$ and the temperature $T$ on the Charm quark decoupling hyper-surface. In Fig.~\ref{fig:ExPlot} the width of the fluctuation of net Charm quark number is plotted over $\kappa\sigma^2$, where we also included lines of constant temperature (solid) and lines of constant volume (dashed). To provide an estimate for the uncertainty, we include the errors of the pQCD calculation as bands. The relevant volume range can be estimated from the analysis in \cite{Andronic:2009qf}, which finds a final freeze-out volume on the order of 5300 fm$^3$. Based on the measured fluctuation data for different experimental values of $\sqrt{\langle(\delta N_{c-\bar c})^2\rangle}$ and $\kappa\sigma^2$ the figure then allows to estimate the Charm quark equilibration temperature and the corresponding volume. 

Additional information and constraints may be obtained from the combined measurement with other susceptibilities.

\section{Conclusion}
We have proposed a method to extract the Charm quark thermalization temperature and volume in the early QGP stage, via the combined application of lattice or perturbative QCD data and experimental measurements.
This method can serve to explore the proposed flavor hierarchy \cite{Bellwied:2013cta} in the thermalization processes and can be used as a thermometer for the  the early stage of the fireball evolution. Our perturbative investigations of the Charm susceptibilities reproduce the trend of the lattice data and the mass effects therein.

\section{Acknowledgments}
  We would like to thank P. Petreczky and R. Bellwied for discussions. This work was supported by the Hessian LOEWE initiative through the Helmholtz International Center for FAIR (HIC for FAIR). It was also supported by DAAD, GSI and BMBF. This work was partially supported by the COST Action THOR, CA15213. TG was supported by the Service pour la science et la technologie pr\`{e}s l'Ambassade de France en Allemagne and Campus France. CH acknowledges support from Suranaree University of Technology (SUT) and SUT-CHE-NRU (FtR.15/2559) project. 

\bibliography{charm_fluc_2016_08_07_TG.bib}

\end{document}